\documentclass{aa}

\usepackage[varg]{txfonts}   
\usepackage[english]{babel}
\usepackage{graphicx}
\usepackage{multicol}
\usepackage{paralist}
\usepackage{setspace}
\usepackage{sidecap}
\usepackage{palatino}
\usepackage{fancyvrb}
\usepackage{color}
\usepackage[T1]{fontenc}
\usepackage[utf8]{inputenc}
\usepackage{amsmath}
\usepackage{amssymb}
\usepackage{amsfonts}
\usepackage{setspace}
\usepackage{listings}
\usepackage{slashed}
\usepackage[nice]{units}
\usepackage{multirow}
\usepackage{rotating}
\usepackage{natbib}
\bibpunct{(}{)}{;}{a}{}{,}

\usepackage[switch]{lineno}

\newcommand{\xmm}{\textit{XMM-Newton}}
\newcommand{\fermi}{\textit{Fermi}/LAT}
\newcommand{\swift}{\textit{Swift}}
\newcommand{\pks}{\object{\mbox{PKS\,2004$-$447}}}
\newcommand{\pmn}{\object{\mbox{PMN\,J0948$+$0022}}}
\newcommand{\pkss}{\object{\mbox{PKS\,1502$+$036}}}
\newcommand{\h}{\object{\mbox{1H\,0323$+$342}}}
\newcommand{\sbs}{\object{\mbox{SBS\,0846$+$513}}}
\newcommand{\fbqs}{\object{\mbox{FBQS\,J1644+2619}}}
\newcommand{\hb}{\object{\mbox{[HB89]\,1219+044}}}
\newcommand{\rlnls}{\mbox{RL-NLS1}}
\newcommand{\gnls}{\mbox{$\gamma$-NLS1}}
\newcommand{\nls}{\mbox{NLS1}}
\newcommand{\oshlack}{\mbox{Osh01}}
\newcommand{\gallo}{\mbox{G06}}

\renewcommand*\apjl{ApJL}

\title{The Gamma-Ray Emitting Radio-Loud Narrow-Line Seyfert 1 Galaxy \pks{}}
\titlerunning{The Gamma-Ray Emitting Radio-Loud Narrow-Line Seyfert 1 Galaxy \pks{} II}
\subtitle{II. The Radio View}
\author{R. Schulz\inst{\ref{inst1},\ref{inst2}}
	\and A. Kreikenbohm\inst{\ref{inst1}, \ref{inst2}}
	\and M. Kadler \inst{\ref{inst1}}
	\and R. Ojha\inst{\ref{inst3},\ref{inst4},\ref{inst5}}
	\and E. Ros \inst{\ref{inst6},\ref{inst7},\ref{inst8}}	
	\and J. Stevens\inst{\ref{inst9}}
	\and P. G. Edwards\inst{\ref{inst9}}
	\and B. Carpenter\inst{\ref{inst3}, \ref{inst5}}
	\and D. Els\"{a}sser\inst{\ref{inst1}}
	\and N. Gehrels\inst{\textbf{\ref{inst3}}}
	\and C. Gro{\ss}berger\inst{\ref{inst10},\ref{inst2}}
	\and H. Hase\inst{\ref{inst11}}
	\and S. Horiuchi\inst{\ref{inst12}}
	\and J.~E.~J. Lovell\inst{\ref{inst13}}
	\and K. Mannheim\inst{\ref{inst1}}	
	\and A. Markowitz\inst{\ref{inst2},\ref{inst14},\ref{inst15}}
	\and C. M\"{u}ller\inst{\ref{inst16}, \ref{inst2}}
	\and C. Phillips\inst{\ref{inst9}}
	\and C. Pl\"otz\inst{\ref{inst11}}
	\and J. Quick\inst{\ref{inst17}}
	\and J. Tr\"{u}stedt\inst{\ref{inst1}}
	\and A.~K. Tzioumis\inst{\ref{inst9}}
	\and J. Wilms \inst{\ref{inst2}}
}
\institute{Lehrstuhl f\"{u}r Astronomie, Universit\"{a}t W\"{u}rzburg, Campus Hubland Nord, Emil-Fischer-Str. 31, 97074 W\"{u}rzburg, Germany \label{inst1}
\and Dr. Remeis-Sternwarte \& ECAP, Universit\"{a}t Erlangen-N\"{u}rnberg, Sternwartstr. 7, 96049 Bamberg, Germany \label{inst2}
\and NASA, Goddard Space Flight Center, 8800 Greenbelt Rd, Greenbelt, MD 20771, USA \label{inst3}
\and University of Maryland, Baltimore County, 1000 Hilltop Cir, Baltimore, MD 21250, USA \label{inst4}
\and The Catholic University of America, 620 Michigan Ave NE, Washington, DC 20064, USA \label{inst5}
\and Max-Planck-Institut f\"{u}r Radioastronomie, Auf dem Hügel 69, 53121 Bonn, Germany \label{inst6}
\and Observatori Astron\`{o}mic, Univ. Val\`{e}ncia, 46980 Paterna Val\`{e}ncia, Spain \label{inst7}
\and Dept. Astronomia i Astrof\`{i}sica, Univ. Val\`{e}ncia, 46100 Burjassot, Val\`{e}ncia, Spain \label{inst8}
\and CSIRO, Astronomy and Space Science, ATNF, PO Box 76 Epping, NSW 1710, Australia \label{inst9}
\and Max-Planck-Institut f\"ur extraterrestrische Physik, Giessenbachstrasse 1, 85741 Garching, Germany \label{inst10}
\and Bundesamt f\"ur Kartographie und Geod\"asie, 93444 Bad K\"otzingen, Germany \label{inst11}
\and CSIRO Astronomy and Space Science, Canberra Deep Space Communications Complex, PO Box 1035, Tuggeranong ACT 2901, Australia\label{inst12}
\and School of Mathematics \& Physics, University of Tasmania, Private Bag 37, Hobart, 7001 Tasmania, Australia\label{inst13}
\and University of California, San Diego, CASS, 9500 Gilman Dr., MC 0424, La Jolla, CA 92093-0424, USA\label{inst14}
\and Alexander von Humboldt Fellow\label{inst15}
\and Department of Astrophysics/MAPP, Radboud University Nijmegen, PO Box 9010, 6500 GL, Nijmegen, The Netherlands\label{inst16}
\and Hartebeesthoek Radio Astronomy Observatory, Krugersdorp, South Africa\label{inst17}
}
\date{Received 21 September 2015 / Accepted 25 October 2015} 

\abstract
{$\Gamma$-ray detected radio-loud narrow-line Seyfert 1 (\gnls) galaxies constitute a small but interesting sample of the $\gamma$-ray loud AGN. The radio-loudest \gnls{} known, \pks, is located in the southern hemisphere and is monitored in the radio regime by the multiwavelength monitoring program TANAMI.}
{We aim for the first detailed study of the radio morphology and long-term radio spectral evolution of \pks, which are essential to understand the diversity of the radio properties of \gnls{s}.}
{The TANAMI VLBI monitoring program uses the Australian Long Baseline Array (LBA) and telescopes in Antarctica, Chile, New Zealand, and South Africa to monitor the jets of radio-loud active galaxies in the southern hemisphere. Lower resolution radio flux density measurements at multiple radio frequencies over four years of observations were obtained with the Australia Telescope Compact Array (ATCA).}
{The TANAMI VLBI image at 8.4\,GHz shows an extended one-sided jet with a dominant compact VLBI core. Its brightness temperature is consistent with equipartition, but it is an order of magnitude below other \gnls{s} with the sample value varying over two orders of magnitude. We find a compact morphology with a projected large-scale size $<11\mathrm{\,kpc}$ and a persistent steep radio spectrum with moderate flux-density variability.}
{\pks{} appears to be a unique member of the \gnls\ sample. It exhibits blazar-like features, such as a flat featureless X-ray spectrum and a core dominated, one-sided parsec-scale jet with indications for relativistic beaming. However, the data also reveal properties atypical for blazars, such as a radio spectrum and large-scale size consistent with Compact-Steep-Spectrum (CSS) objects, which are usually associated with young radio sources. These characteristics are unique among all \gnls{s} and extremely rare among $\gamma$-ray loud AGN.}

\begin{document}

\maketitle

\section{Introduction}
\label{sec-Intro}
 
	The radio-loud versions of narrow-line Seyfert 1 (\rlnls) galaxies have attracted growing interest in recent years, especially since \cite{Abdo2009c} reported the detection of $\gamma$-ray emission from \pmn{} with the LAT instrument on-board the \textit{Fermi} $\gamma$-ray satellite \citep{Atwood2009}. 
	\nls{} galaxies harbour an active galactic nucleus (AGN) which exhibits both broad and narrow emission lines as in typical type-1 AGN. The broad lines, however, are substantially narrower than in other type-1 AGN. \nls{} galaxies show strong and narrow H$\beta$ emission with a flux ratio of \ion{O}[III]/H$\beta\le3$, a full width at half maximum of broad H$\beta_\mathrm{FWHM}\le2000\,\mathrm{km\,s}^{-1}$ and strong Fe\,II emission \citep{Osterbrock1985}. It is not uncommon for \nls{s} to show variability in flux and photon index (e.g., \citealt{Boller1997}). 
	Like quasars, they can be divided into radio-quiet and radio-loud categories, where the radio-loudness $\mathrm{RL}_\nu$ is defined as the ratio of the radio flux density at frequency $\nu$ to the optical flux at 440\,nm. A source is commonly characterized as radio-loud when $\mathrm{RL}_\mathrm{1.4GHz}>10$. 
	
	\rlnls\ galaxies are rare among radio-loud AGN \citep{Komossa2006,Zhou2006}.
	However, the total number of known \rlnls{s} has increased in recent years. The first dedicated study and search of \rlnls{s} was performed by \cite{Komossa2006}, yielding a sample of 11 sources most which were previously not known to be \rlnls{s}. Then, \cite{Yuan2008} compiled a genuine sample of 23 \rlnls{s} with $\mathrm{RL}_{1.4\mathrm{GHz}}>100$, with black hole masses of $10^6$--$10^8\,M_\sun$ and Eddingtion ratios $R_{\mathrm{Edd}}\approx 1$. Based on these and other studies, \cite{Foschini2015} and \cite{Berton2015} have a combined sample of 60 sources with $\mathrm{RL}_{5\mathrm{GHz}}>10$. Only a small number of \rlnls{s} have been detected by \fermi{} (\gnls, \citealt{Abdo2009b,Foschini2015,Dammando2015,Yao2015}, see Sect. \ref{sub-sec:comparison}) with most of them listed in the Third \fermi{} Catalog of AGNs with $b < -10\degr$ and $b > 10\degr$ (3LAC, \citealt{Ackermann2015}). \gnls{} galaxies belong to the exclusive group of so-called non-blazar AGN in the 3LAC which make up only about $\sim 2\,\%$ of the 3LAC sources. The majority of sources in the 3LAC are blazars, which appear to be hosted by elliptical galaxies with central black hole masses above $10^8\,M_\sun$ (e.g., \citealt{Marscher2009} and references therein). They have powerful jets aligned close to the line of sight. 
	
	\pks{} is a \rlnls\ at a redshift of $z=0.24$ \citep{Drinkwater1997}. It stands out of \gnls{s} because of its unusual radio properties. With $\mathrm{RL}_{4.85\mathrm{GHz}}$=1700--6300, depending on the optical flux (\citealt{Oshlack2001}, hereafter \oshlack), it has the highest radio-loudness and shows a steep radio spectrum with a spectral index\footnote{The spectral index $\alpha$ is defined as $S_\nu\propto \nu^\alpha$, where $S_\nu$ is the flux density at frequency $\nu$.} of $\alpha_r<-0.5$ (\oshlack; \citealt{Gallo2006}, hereafter \gallo). 
	Observations with the Australia Telescope Compact Array (ATCA) show an unresolved source, suggesting it can be classified as a Compact Steep-Spectrum (CSS) radio source (\oshlack, \gallo). 
	
	\pks{} exhibits only weak Fe\,II emission, with $EW_{\mathrm{Fe\,II}}\le10\AA$, a flux ratio \ion{O}[III]/H$\beta=1.6$ and H$\beta_\mathrm{FWHM}=1447\,\mathrm{km\,s}^{-1}$ (\oshlack).  As a result, its \nls{} classification has been under discussion (e.g., \citealt{Zhou2006,Yuan2008}). Despite its steep spectral index, it was included in the sample of flat spectrum \rlnls\ studied by \cite{Foschini2015}.
	
	This is the second in a series of papers presenting results from broadband observations of \pks. The observations reported in this series were conducted as part of the TANAMI\footnote{Tracking Active Nuclei with Austral Milliarcsecond Interferometry \url{http://pulsar.sternwarte.uni-erlangen.de/tanami/projects/}} multiwavelength program \citep{Ojha2010,Kadler2015} that monitors $\gamma$-ray loud AGNs south of $-30^{\circ}$ declination. In a separate paper (\citealt{Kreikenbohm2014}; hereafter Paper I) we discussed X-ray observations with \xmm{} and \swift{}, finding a flat blazar-like power law and moderate variability on time scales down to months with an observed luminosity of (0.7--2.6)$\times 10^{44}$ erg/s. In this paper, we present the first Very-Long-Baseline Interferometry (VLBI) image for \pks{} at 8.4\,GHz (the highest radio frequency VLBI image of the source in the literature so far) as well as the first long-term simultaneous multi-frequency radio monitoring with ATCA.
	
	In the next section, we will discuss the data reduction followed by a presentation of our results in Sect. \ref{sec-results}. In Sect. \ref{sec-discussion}, we will discuss our findings and compare them within the sample of \gnls. 
	
	We adopt a $\Lambda$CDM cosmology, with \mbox{$H_0=70\,\mathrm{km\,s}^{-1}\,\mathrm{Mpc}$}, $\Omega_M=0.3$, and $\Lambda=0.7$ \citep{Freedman2001}.

\section{Observations and data reduction}
\label{sec-obs}

	\subsection{ATCA observations}
	\label{sec-obs-atca}
		As part of the TANAMI program, ATCA has been monitoring $\gamma$-ray detected AGN at frequencies between $\sim$ 5\,GHz and 40\,GHz since 2007 \citep{Stevens2012}. \pks{} has been included in the program since 2011 May 17. Each observing frequency is the centre of a $2\mathrm{\,GHz}$ wide band. PKS\,1934$-$638 has been used as a flux density calibrators. Table \ref{tab-AtcaObsInfo} lists the data obtained until 2014 Jun 13 with statistical errors. In addition, we make use of monitoring data from the ATCA Calibrator Database\footnote{\url{http://www.narrabri.atnf.csiro.au/calibrators/}} extending the spectral coverage at selected epochs to 1.7\,GHz and 45\,GHz.
		
		ATCA observations of this source show a moderate defect\footnote{Defect is defined as: $([S_\mathrm{sca}/S_\mathrm{vec}] - 1)\times100\%$, where $S_\mathrm{sca}$ is the scalar-averaged flux density, and $S_\mathrm{vec}$ is the vector-averaged flux density.} at 17\,GHz and above and very low closure phase. The closure phase is an interferometric quantity which is independent of errors introduced at any individual array element, and will always be zero (to within the noise of the measurement) if an unresolved source is the dominant source of the flux density in the field of view. The closure phase will deviate from zero if there are other sources producing significant flux density within the field.
		
		The defect uses the difference between the scalar and vector averaged flux densities to give us more information about the field. Vector averaging uses the phase information obtained by the interferometer to attenuate sources of flux density away from the phase centre of the measurement. The effect of this averaging is to make a measurement only of the flux density of the source that the array is pointing at directly, and this effect is strongest when using widely spaced array elements. Scalar averaging however tends to capture a lot more of the flux density being seen by each array element, regardless of where the flux density is being produced within the field of view.
		
		For a field of view which is empty apart from a point source at the phase centre, both the closure phase and defect would be zero, to within the noise level of the measurement. A significant defect, and a closure phase close to zero would suggest that some amount of flux density is coming from sources that are not at the phase centre - what we call ``confusing structure''. Such confusing structure could affect the calibration, and as this effect is frequency dependent, it will tend to make the radio spectrum slightly steeper. However, we consider this to be negligible for our ATCA data. The defect at a frequency of 17\,GHz and above indicates that the source is resolved by ATCA. Conversely, the lack of a defect at 9.0\,GHz shows that the source is unresolved for the ATCA at this frequency. In combination, the maximum scales of extended emission seem to range between $1\farcs7$ and $0\farcs8$.

	\subsection{VLBI and VLA observations}
	\label{sec-obs-tanami}
	
		\pks{} was first observed in the framework of the TANAMI VLBI program on 2010 Oct 29. This observation was made at a frequency of 8.4\,GHz. The TANAMI array consists of the five telescopes comprising the Australian Long Baseline Array (LBA),  in combination with telescopes in Antarctica, Chile and South Africa (see Table \ref{tab-TanamiObsInfo}). The data were correlated on the DiFX software correlator \citep{Deller2007, Deller2011} at Curtin University in Perth, Western Australia. The data were then calibrated and imaged following \cite{Ojha2010}. Following \cite{Ojha2010}, we assume conservative uncertainties of the amplitude calibration for the TANAMI data of 20\%.
		
		We also analyzed the only other available, archival VLBI observation obtained by the Very Long Baseline Array (VLBA) on 1998 Oct 12 (project code: BD0050). We calibrated the data using the Astrophysical Image Processing System (\textsc{AIPS}, \citealt{Greisen2003}). Hybrid imaging was conducted with the software package \textsc{Difmap} \citep{Shepherd1994}. We used uniform weighting to achieve the highest angular resolution. \cite{Lister2005} estimate an uncertainty of 5\% for their 15\,GHz VLBA data. Given the very low declination of \pks{} for the VLBA and after repeating the imaging process several times, we assume a conservative uncertainty of 10\% for the amplitude calibration of the VLBA data.
		
		The Very Large Array (VLA) participated in `B' configuration in this VLBA observation. We extracted the VLA data from the archive and calibrated and imaged the data independently to check for missing flux in the VLBA image.
		
		The parameters of the various images are given in Table~\ref{tab-TanamiObsInfo}. The noise level of the image $\sigma_\mathrm{RMS}$ was determined by fitting the noise-dominated pixels with a Gaussian distribution \citep{Boeck2012} with the Interactive Spectral Interpretation System (\textsc{ISIS}) package \citep{Houck2000}.
		
		\begin{table*}
			\caption{Details of interferometric observations and image parameters}
			\label{tab-TanamiObsInfo}
			\centering
			\begin{tabular}{@{}l@{\;\;}c@{\;\;}p{2.5cm}@{\;\;}c@{\;\;}c@{\;\;}c@{\;\;}c@{\;\;}c@{\;\;}c@{\;\;}c@{}}
				\hline \hline	
				Date & Freq. & Array\tablefootmark{a} & Taper & $S_\mathrm{peak}$ & $\sigma_\mathrm{RMS}$ & $S_\mathrm{total}$ & $b_\mathrm{maj}$\tablefootmark{b} & $b_\mathrm{min}$\tablefootmark{b} & P.A.\tablefootmark{b}\\
				$[\mathrm{yyyy}$-$\mathrm{mm}$-$\mathrm{dd}]$ & [GHz] & & [M$\lambda$] & [mJy\,beam$^{-1}$] & [mJy\,beam$^{-1}$] & [mJy] & [mas] & [mas] & [\degr] \\
				\hline
				2010-10-29 & 8.4 & PKS-AT-MP-HO-CD-HH & - & $195\pm 39$ & 0.079 & $294\pm 59$ & 2.83 & 0.62 & $-2.7$ \\
				&	&	& 60 & $249\pm 50$ & 0.064 & $301\pm 60$ & 5.61 & 4.42 & $-63$ \\
				1998-10-12 & 1.5 & VLBA & - & $213\pm 21$ & 0.41 & $605\pm 61$ & 14.1 & 4.16 & $-9.0$\\
				1998-10-13 & 1.5 & VLA & - & $726\pm 73$ & 0.16 & $730\pm 73$ & $1.15\times 10^4$ & $3.0\times 10^3$ & $-1.7$\\
				\hline
			\end{tabular}
			\tablefoot{
				\tablefoottext{a} {PKS: Parkes (64\,m), AT: ATCA (5$\times$22\,m), MP: Mopra (22\,m), HO: Hobart (26\,m), CD: Ceduna (30\,m), HH: Hartebeesthoek (26\,m); VLBA 8$\times$25m (no Hancock and Brewster), including phased VLA antenna Y27; VLA in B configuration with 27$\times$25\,m stations.}
				\tablefoottext{b} {Major and minor axis and position angle of the restoring beam.}
			}
		\end{table*}

\section{Results}
\label{sec-results}

	\subsection{ATCA monitoring}
	\label{subsec-atca}
	
\begin{figure}
			\centering
			\includegraphics[width=\linewidth]{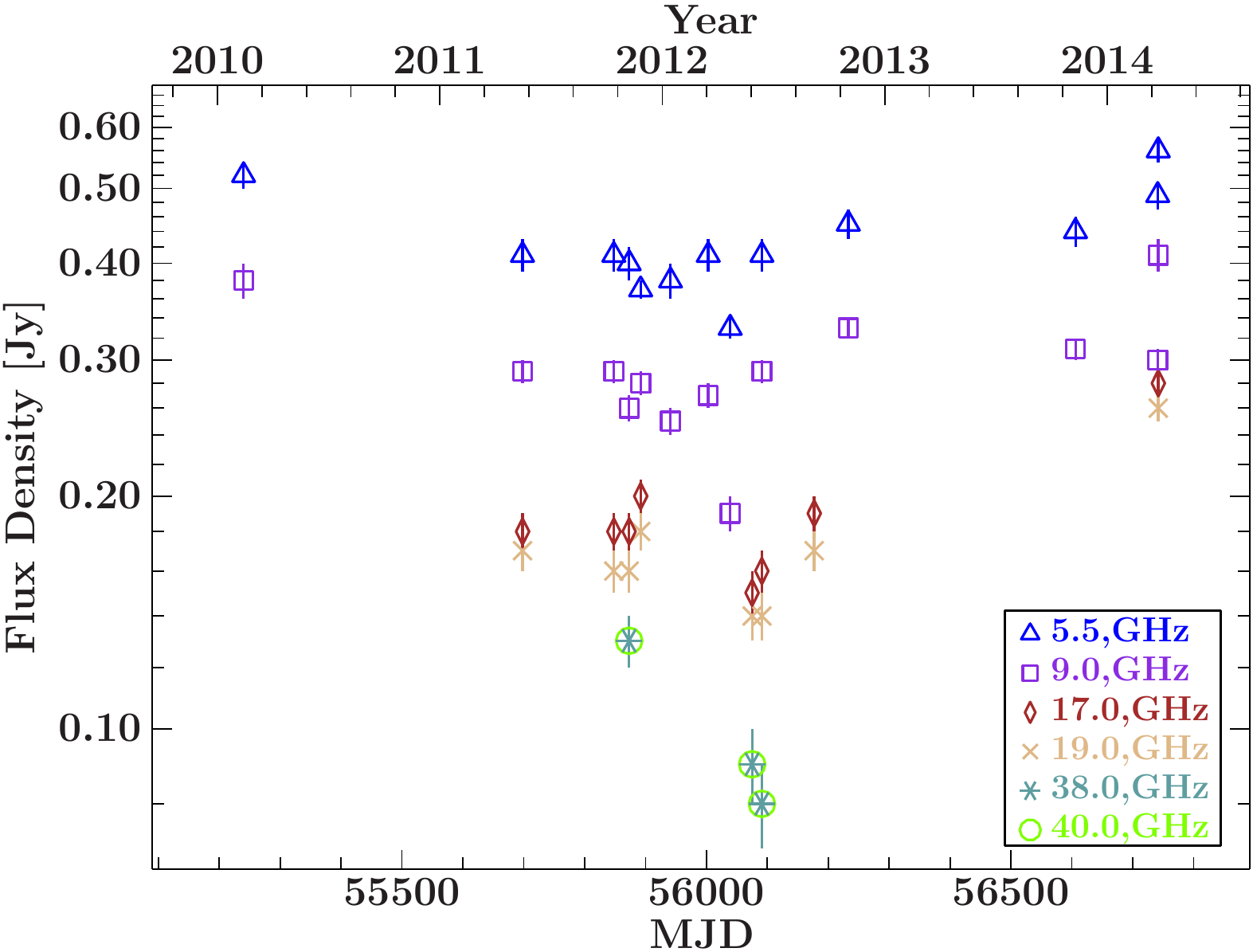} 
			\caption{ATCA radio light curve at 5.5\,GHz (triangles), 9.0\,GHz (rectangles), 17.0\,GHz (diamonds), 19.0\,GHz (crosses), 38.0\,GHz (asterisks) and 40\,GHz (circles) using only frequencies with more than two data points (see Table~\ref{tab-AtcaObsInfo}).}
			\label{fig-atca-lc}
		\end{figure}

\begin{figure}
			\centering
			\includegraphics[width=\linewidth]{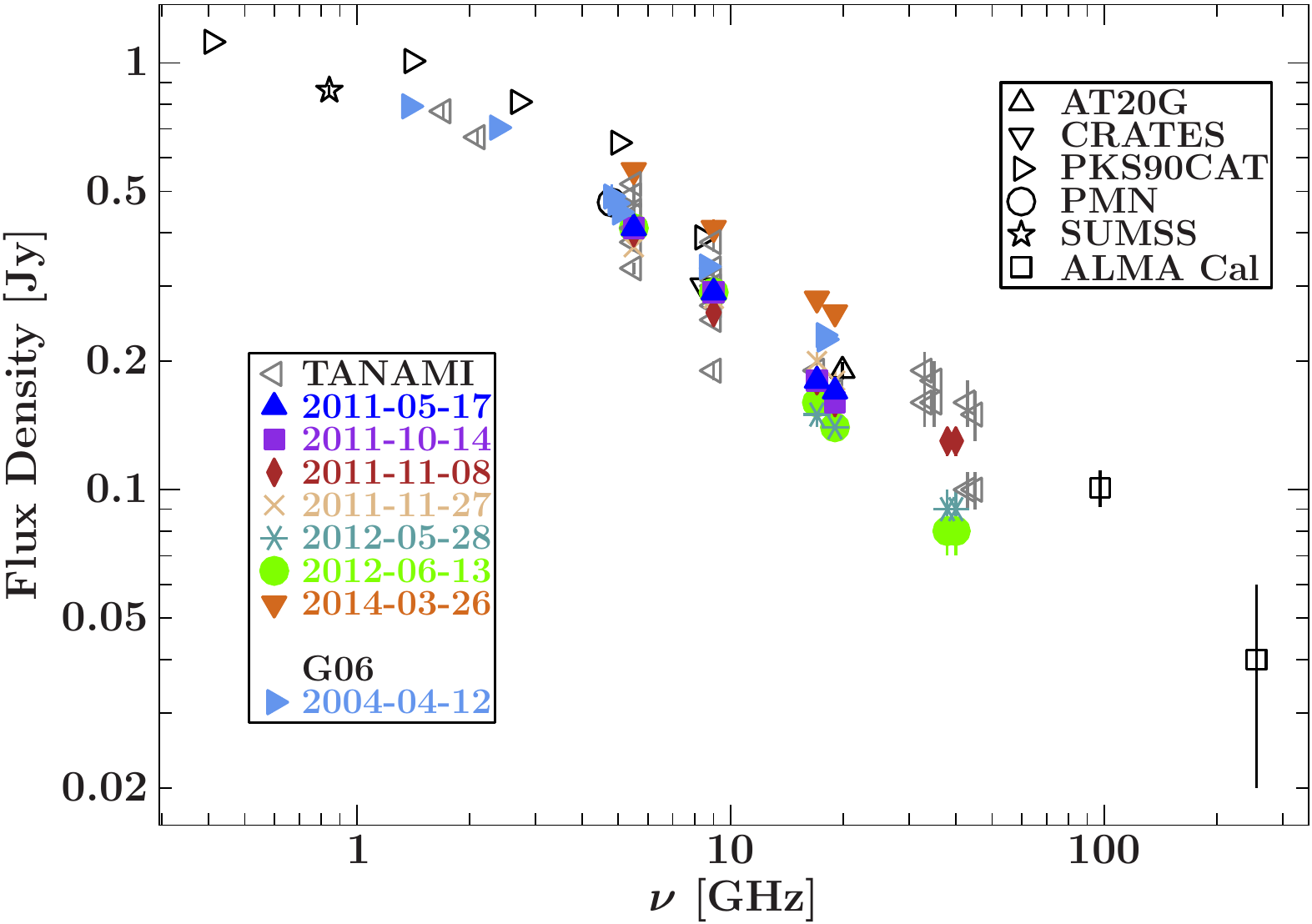} 
			\caption{Spectrum of ATCA monitoring (Fig. \ref{fig-atca-lc}), between 1.7\,GHz and 45\,GHz from 2010 Feb 13 to 2014 Mar 26. Data for which spectral index fits were performed, are shown in colour. Simultaneous archival data taken from \gallo\ (blue, right-oriented triangles). Non-simultaneous archival data from various catalogues are shown in gray: AT20G (asterisk, \citealt{Murphy2010,Chhetri2013}), CRATES (bottom-oriented triangles, \citealt{Healey2007}, PKS90CAT (right-oriented triangles, \citealt{Wright1990}), PMN (circles, \citealt{Griffith1994,Wright1994}), SUMSS (stars, \citealt{Mauch2013}), and ALMA Cal (rectangles, \citealt{Fomalont2014}).}
			\label{fig-atca-spectrum}
		\end{figure}

		Figure \ref{fig-atca-lc} shows the ATCA light curve between 5.5\,GHz and 40\,GHz for the period 2010 Feb 13 through 2014 Mar 26. Figure \ref{fig-atca-spectrum} depicts the corresponding radio spectrum supplemented by individual ATCA calibrator database measurements at 1.7\,GHz, 2.1\,GHz, 33.0\,GHz, 35.0\,GHz, 43.0\,GHz, and 45.0\,GHz. This marks the first time that several simultaneous multi-frequency radio observations have been obtained for \pks.
		
		The data show only moderate variability in flux density up to a factor of two and a persistent steep radio spectrum over four years (see Table~\ref{tab-AtcaSpecFit}). There are two possible states of increased activity. First, the 38/40\,GHz data indicate a possible weak flare on 2011 Nov 08 without any obvious counterparts at lower frequencies. Second, the flux density increased at 5.5\,GHz and 9.0\,GHz on 2014 Mar 26. The relatively sparse sampling makes it difficult to study the evolution of the light curve in detail. 
		
		We adopt the variability index $V_\nu$ at frequency $\nu$ \citep{Hovatta2008} as a basic quantity for the strength of the flaring activity,
		\begin{align}
		V_\nu = \frac{(S_{\nu,\mathrm{max}} - \sigma_{\nu,\mathrm{max}}) - (S_{\nu,\mathrm{min}} + \sigma_{\nu,\mathrm{min}})}{(S_{\nu,\mathrm{max}} - \sigma_{\nu,\mathrm{max}}) + (S_{\nu,\mathrm{min}} + \sigma_{\nu,\mathrm{min}})}
		\label{eq:varindex}
		\end{align}
		where $S_{\nu,\mathrm{max}}$ and $S_{\nu,\mathrm{min}}$ are the maximum and minimum flux density at frequency $\nu$ and as a basic indicator $\sigma_{\nu,\mathrm{max}}$ and $\sigma_{\nu,\mathrm{min}}$ the corresponding uncertainties. More sophisticated variability quantities would require better sampling (see e.g., \citealt{Richards2011} for an overview). We calculate $V_\nu$ between 5.5\,GHz and 19.0\,GHz which are the frequencies with the best coverage. The results indicate similar moderate variability between 0.22 and 0.34, at all four frequencies.
		
		We investigate the simultaneous ATCA spectra by fitting a power-law $S_\nu=k\nu^\alpha$ where $k$ is a proportionality constant and $\alpha$ is the spectral index. We use a linear regression approach with $\log S_\nu = \log k + \alpha\log(\nu/\mathrm{GHz})$, where $\nu$ is in units of GHz to determine the spectral index and its uncertainty. Where possible, we produced fits for different spectral ranges, which are listed in Table~\ref{tab-AtcaSpecFit}. Above 5.0\,GHz, the spectral index varies moderately between $\sim -0.5$ and $-0.9$.We have inspected the ratio of the data and the model and find that a power-law fit is sufficient to describe the analysed spectra. Only in the case of the data from 2011 Nov. 08, the power-law fit between 5.5\,GHz and 40\,GHz fits only marginally well due to a flattening of the spectrum above 17\,GHz. (see also Sect. \ref{subsec-Discussion-radio-spectrum}.)
		
		\begin{table}
			\caption{Results of power-law fit to various spectral ranges of the ATCA data ($S_\nu=k\nu^\alpha$)}
			\label{tab-AtcaSpecFit}
			\centering
			\begin{tabular}{@{}lccc@{}}
				\hline \hline
				Date & $\alpha_{5.5-40}$\tablefootmark{a} & $\alpha_{5.5-19}$\tablefootmark{b} &  $\alpha_{17-40}$\tablefootmark{c} \\
				\hline
				2004-04-12\tablefootmark{d} & 	& $-0.56\pm0.03$ & \\
				\hline
				2011-05-17 & 	& $-0.72\pm 0.05$ & \\
				2011-10-14 & 	& $-0.75\pm0.05$  & \\
				2011-11-08 & $-0.57\pm 0.03$ & $-0.71\pm 0.05$ & $-0.36\pm 0.09$ \\
				2011-11-27 & 	& $-0.56\pm0.04$ &\\
				2012-05-28 & 	&	& $-0.61\pm 0.12$\\
				2012-06-13 & $-0.86\pm 0.04$ & $-0.86\pm0.06$ & $-0.82\pm 0.13$\\
				2014-03-26 &	& $-0.62\pm 0.04$ &\\
				\hline
			\end{tabular}
			\tablefoot{
				\tablefoottext{a} {Parameters of fit to data between 5.5\,GHz and 40\,GHz.}
				\tablefoottext{b} {Parameters of fit to data between 4.8\,GHz and 19\,GHz, i.e., for \gallo{} data: 4.8--18\,GHz, for our ATCA data: 5.5--19\,GHz.}
				\tablefoottext{c} {Parameters of fit to data between 17\,GHz and 40\,GHz.}
				\tablefoottext{d} {Data taken from \gallo.}
			}
		\end{table}

	\subsection{TANAMI VLBI image}
	\label{subsec-tanami}

\begin{figure}
			\centering
			\includegraphics[width=\linewidth]{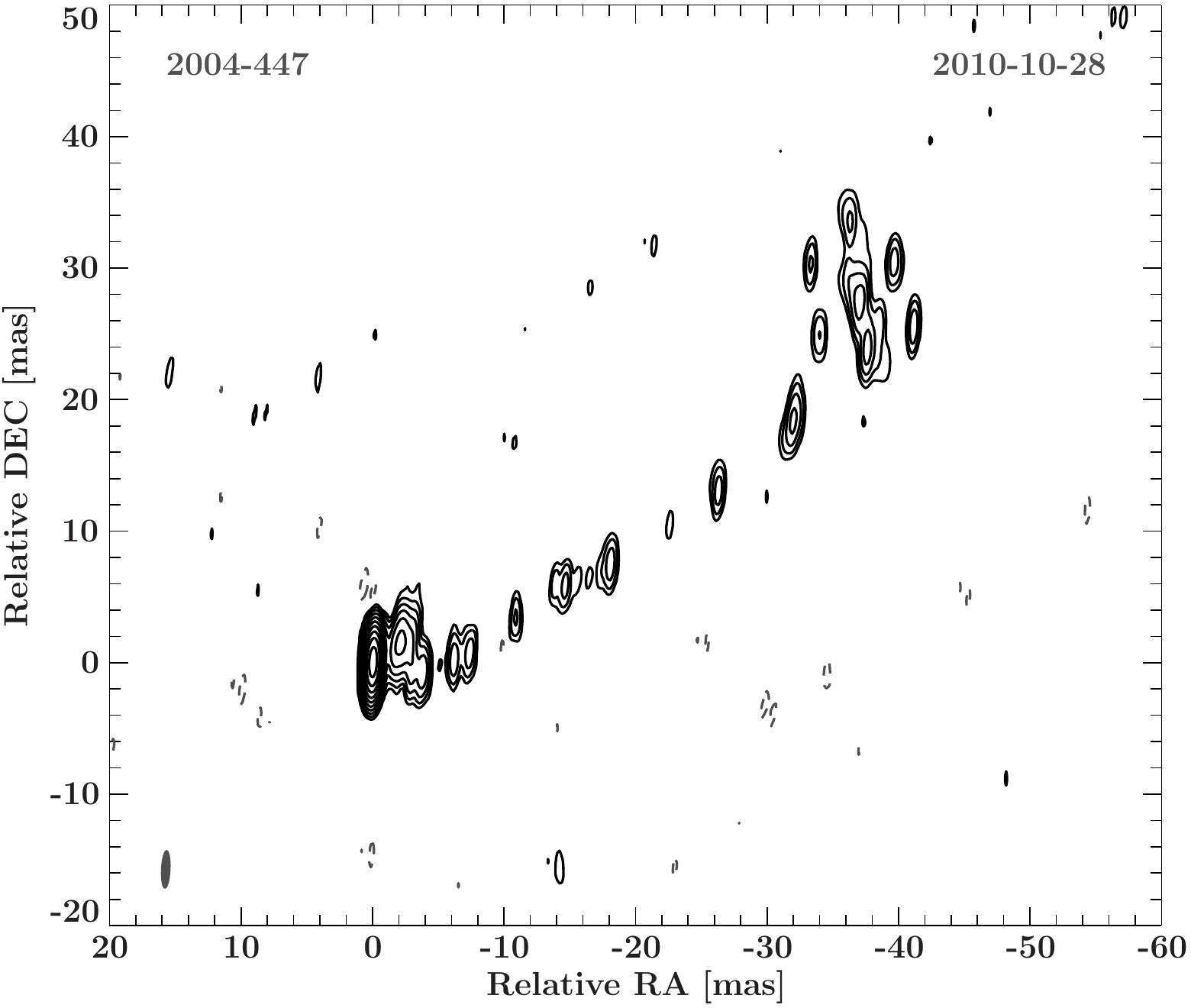}
			\includegraphics[width=\linewidth]{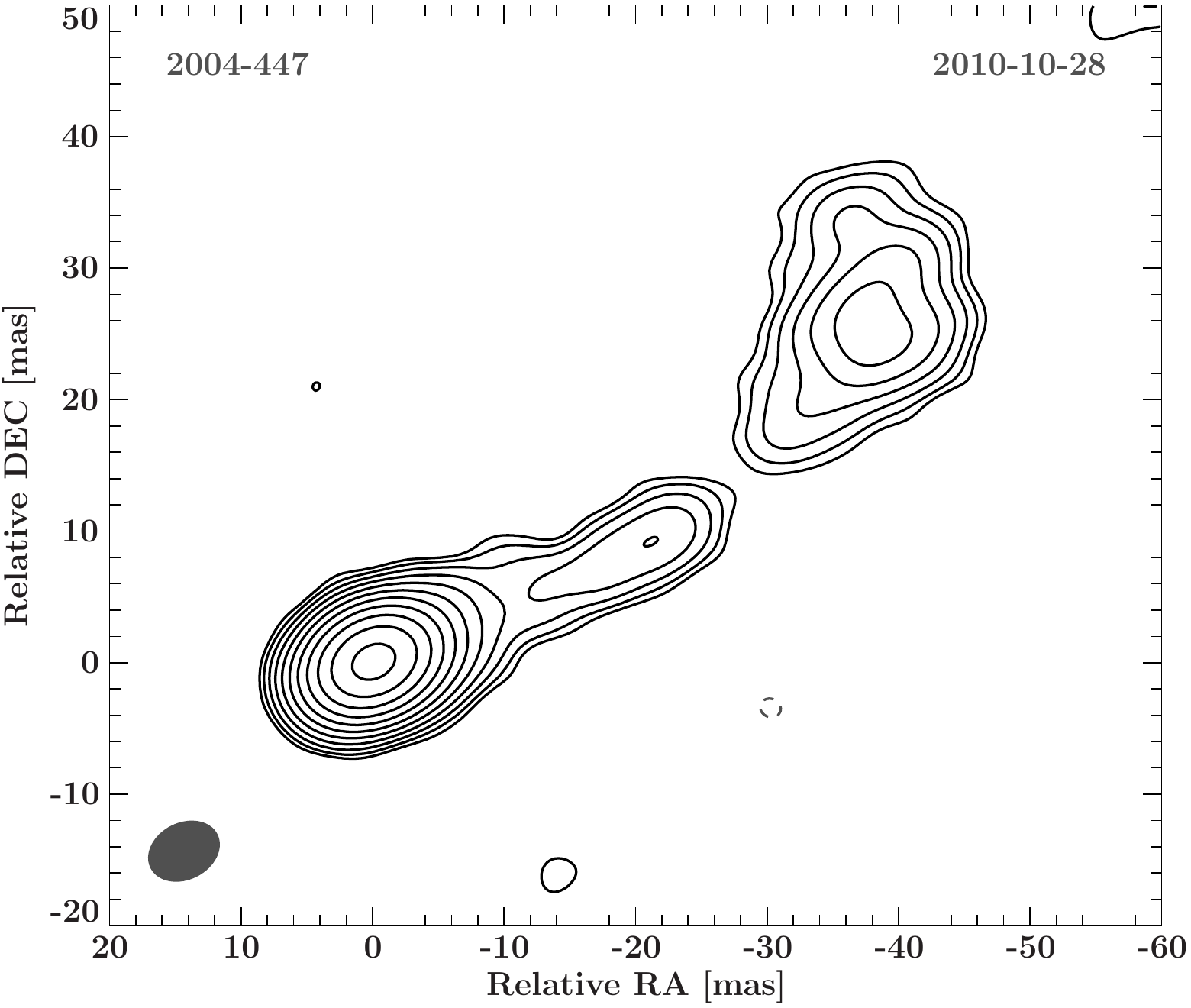} 
			\caption{Top: Full-resolution TANAMI image of 2004-447 at 8.4\,GHz; bottom: Tapered TANAMI image of 2004-447 at 8.4\,GHz. The contour lines of each image start at three times the rms of the image, increasing logarithmically by a factor of two. Negative flux density noise peaks are shown by dashed, grey contour lines. The gray ellipse at the bottom left of both images depicts their respective restoring beams. The image parameters are shown in Table \ref{tab-TanamiObsInfo}.}
			\label{fig-tanami-img}
		\end{figure}

		We present the first TANAMI 8.4\,GHz VLBI image of \pks{} in Fig. \ref{fig-tanami-img}. The brightest feature, which we identify as the VLBI core, is located at the origin of the map. A one-sided jet extends in the north-west direction from the VLBI core. While the jet is partially resolved at full-array resolution beyond 10\,mas from the centre, the tapered image in the bottom panel of Fig. \ref{fig-tanami-img} shows a continuous jet out to $\sim$ 30\,mas and a more diffuse region further downstream with an enhanced emission region or `hot spot' at $\sim$ 45\,mas. This image was obtained by applying a Gaussian taper with a strength of 0.1 at 60\,M$\lambda$ to the visibility data and by repeating the full hybrid imaging process.
		The untapered TANAMI image contains a total flux density of $(294\pm 59)\,\mathrm{mJy}$. The dynamic range $\mathrm{DR}$ of the image, i.e., the ratio between the peak flux density and five times the rms noise is $\mathrm{DR}_{8.4GHz}\approx 490$. 
		
		In order to further characterize the core, we deleted \textsc{CLEAN}-components contained within 1\,mas from the core and fit an elliptical Gaussian component to the core, using \textsc{ISIS} \citep{Houck2000} with a direct link to \textsc{DIFMAP} \citep{Grossberger2014}. This enables us to calculate the given $1\sigma$-uncertainties using $\chi^2$-statistics. The fit parameters are given in Table~\ref{tab-modelfit}. The compact core component yields a flux density of $(229.8^{+0.4}_{-0.2_\mathrm{stat}}\pm35_\mathrm{sys})\mathrm{\,mJy}$, which is almost 80\% of the total flux density. We calculated the brightness temperature $T_b$ following e.g., \cite{Kovalev2005} to be $(5.7\pm 1.4)\times 10^{10}\,\mathrm{K}$. The $1\sigma$ error is predominantly determined by the uncertainty of the axis ratio of the component. The brightness temperature is well below the inverse-Compton limit of $10^{12}\mathrm{\,K}$ \citep{Kellermann1969} and is consistent with the equipartition limit of $\sim 10^{11}\,\mathrm{K}$ \citep{Readhead1994}. Additionally, we fit a circular Gaussian distribution to the `hot spot' by deleting only the \textsc{CLEAN} components within this region and measure a brightness temperature of about $10^7 \mathrm{\,K}$.

	\subsection{VLBA and VLA image}
	\label{subsec-vlba}

\begin{figure}
			\centering
			\includegraphics[width=\linewidth]{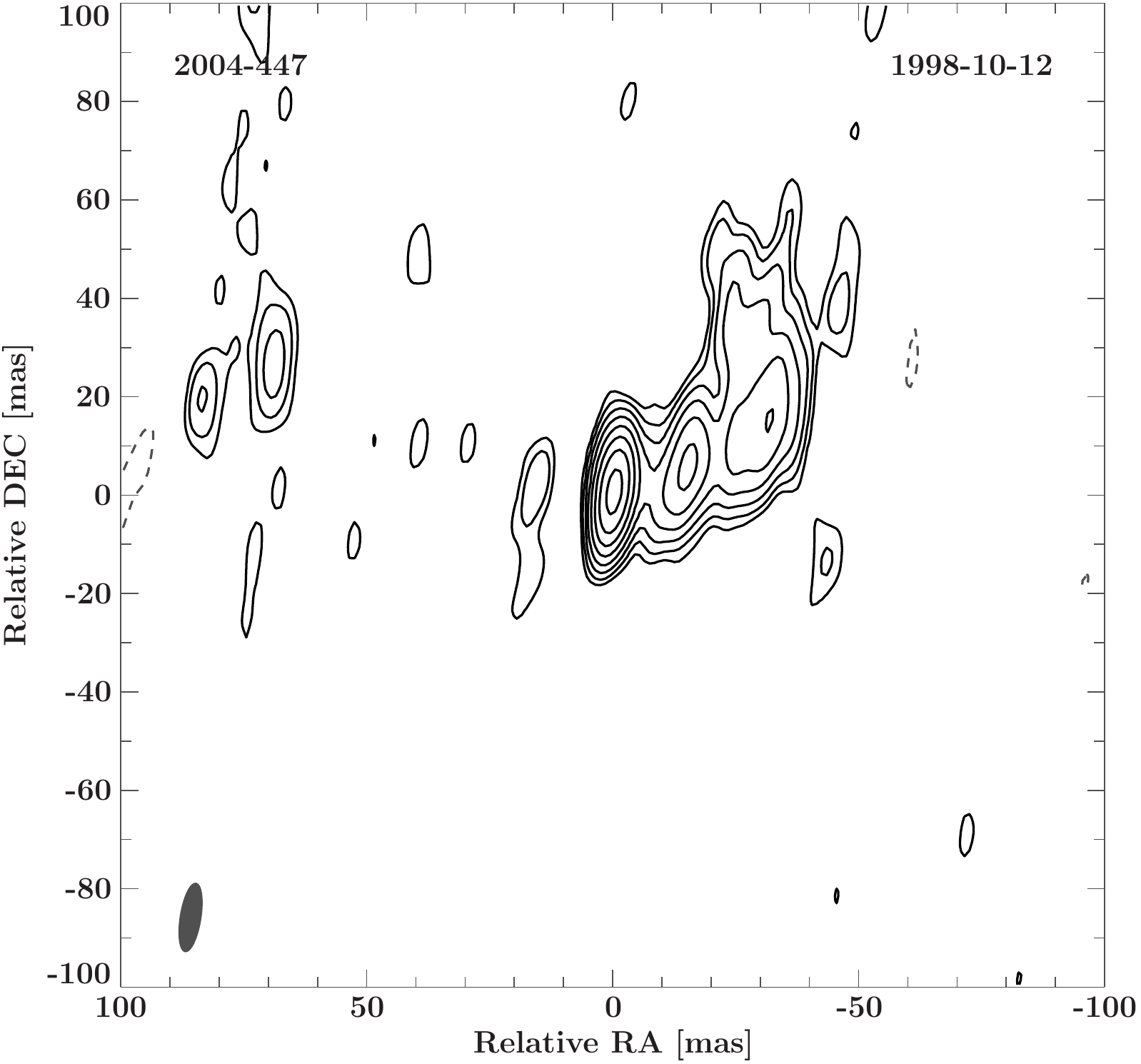} 
			\includegraphics[width=\linewidth]{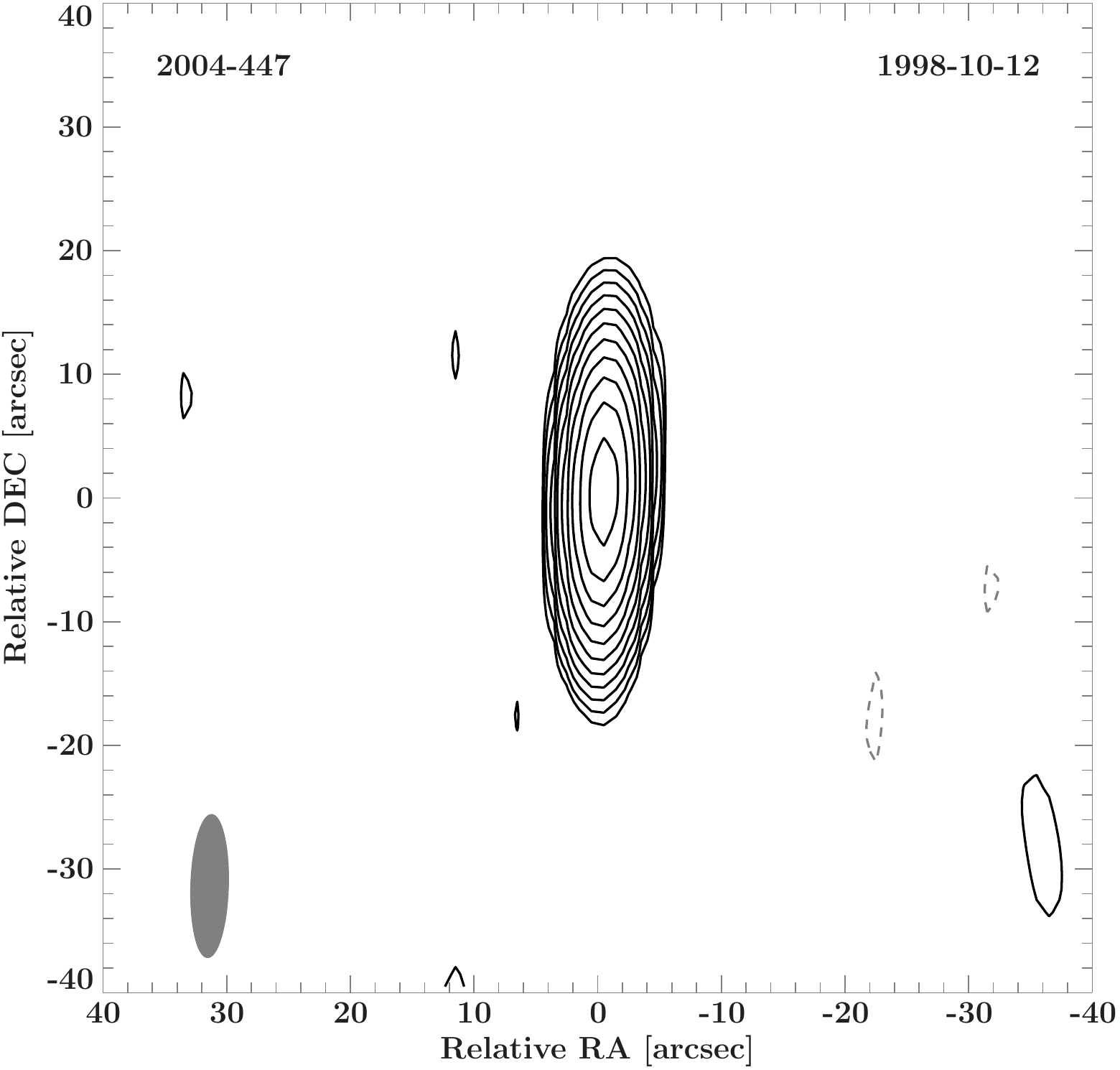} 
			\caption{VLBA (top) and VLA (bottom) image of 2004-447 at 1.5\,GHz with the image parameters given in Tab. \ref{tab-TanamiObsInfo}. Contour lines begin at 3 times the image rms and increase logarithmically by a factor of 2. Grey dashed contour lines indicate negative flux density. The gray ellipse in the bottom left corner depicts the restoring beam.}
			\label{fig-vlba-vla}
		\end{figure}

		Figure \ref{fig-vlba-vla} shows \pks{} at 1.5\,GHz based on the 1998 Oct 12 VLBA observation. We find a total flux density in the image of $(605\pm 61)\mathrm{\,mJy}$ and a dynamic range of $DR_{1.5\mathrm{GHz}}\approx 100$, a comparatively low level due to the very low declination of the source for the VLBA and the resulting sparse distribution of visibilities in the $(u,v)$-plane.
		The radio morphology shows a bright central core with a jet extending to the north-west and substantial diffuse emission on the eastern side with a total extent of about 140\,mas. The `hot spot' at $\sim 45\mathrm{\,mas}$ from the core shows a brightness temperature of $\sim 10^7\mathrm{\,K}$ at 1.5\,GHz.
		
		Following Sect. \ref{subsec-tanami} we fit an elliptical component to the core, accounting for \textsc{CLEAN}-components within 5\,mas from the core. The resulting component has a flux density which contains roughly 40\% of the total flux density (see Table~\ref{tab-modelfit}). We derive a core brightness temperature of $(3.4\pm 0.3)\times 10^{10}\mathrm{\,K}$ at 1.5\,GHz.
		
		The VLA image in Fig. \ref{fig-vlba-vla} shows \pks{} to be compact at 1.5\,GHz for the VLA in B configuration. We fitted an elliptical Gaussian component to the visibility data, which results in an unresolved component sufficient to describe the visibility data without significant residuals. No extended emission is detected using a detection limit of three times the noise level ($0.48\mathrm{\,mJy/beam}$). Hence, the beam of the VLA observation limits the projected linear size to less than $11\mathrm{\,kpc}$. The total flux density in the VLBA and VLA image is consistent considering the uncertainties in amplitude calibration.  The value of the total flux density in the VLBA and VLA image indicates that $\sim 80\%$ of the source flux density is located within $\sim 140\mathrm{\,mas}$ corresponding to $\sim 530\mathrm{\,pc}$ in east-west direction, i.e., in the direction of the minor axis of the VLA beam. 
		
		\begin{table*}
			\caption{Parameters of the elliptical Gaussian component fit to the central feature of the TANAMI and VLBA images}
			\label{tab-modelfit}
			\centering
			\begin{tabular}{cccccc}
				\hline \hline	
				$\nu$ & $S_\mathrm{c}$ & $a_\mathrm{maj}$\tablefootmark{a} & ratio\tablefootmark{a} & $\theta_\mathrm{PA}$\tablefootmark{a} & $T_\mathrm{b}$\tablefootmark{b}\\
				$[$GHz$]$ & $[$mJy$]$ & [mas] & [mas] & [\degr] & [$10^{10}$\,K]\\
				\hline
				8.4 & $229.8^{+0.4}_{-0.2_\mathrm{stat}}\pm35_\mathrm{sys}$  & $0.3651\pm0.0016$ & $0.64^{+0.15}_{-0.14}$ & $88\pm 3$ & $5.7\pm 1.4$\\
				1.5 & $248.8\pm0.4_\mathrm{stat}\pm25_\mathrm{sys}$ & $2.93^{+0.14}_{-0.12}$ & $0.63\pm0.04$ & $31^{+8}_{-7}$ & $3.4\pm 0.3$\\
				\hline
			\end{tabular}
			\tablefoot{Errors are given in $1\sigma$ interval. 
				\tablefoottext{a} {Major axis, ratio of major and minor axis and position angle of the component}
				\tablefoottext{b} {The highest uncertainty of $S_\mathrm{c}$, $a_\mathrm{maj}$ and the axis ratio was taken to calculate the error in the brightness temperature.}
			}
		\end{table*}

\section{Discussion}
\label{sec-discussion}
	
	\subsection{Discussion of the radio spectrum}
	\label{subsec-Discussion-radio-spectrum}
		
		Our ATCA monitoring shows that \pks{} exhibits a persistent steep radio spectrum above 5\,GHz over four years of monitoring. This is consistent with earlier snapshot radio-spectral observations in 2001 and 2004: \oshlack\ reported a spectral index between 2.7\,GHz and 5.0\,GHz of $\alpha_{2.7-5}=-0.67$ from a power-law fit to simultaneous ATCA observations. \gallo\ conducted ATCA observations between 1.4\,GHz and 17.9\,GHz on 2004 Apr 12/13, deriving a spectral index of $\alpha_{1.4-18}=-0.52$.
		
		In order to compare changes in the spectra from our ATCA TANAMI monitoring and \gallo, we select an overlapping spectral range, i.e., between 4.8\,GHz and 19\,GHz. We fit these data points as described in Sect. \ref{subsec-atca}. The results are listed in Table \ref{tab-AtcaSpecFit}. The data show moderate spectral variability with the spectral slope varying between $\sim -0.5$ and $\sim -0.9$ with a weighted mean of $\alpha_\mathrm{wm}=-0.68\pm 0.10$ and a median of $\alpha_\mathrm{median}\approx -0.72$. The mean spectral index is consistent with \oshlack.
		
		A possible flattening above 17\,GHz is observed in epoch 2011 Nov 08. This is due to the aforementioned increased flux density at 38\,GHz and 40\,GHz. Intriguingly, the X-ray light curve (presented in Paper I) shows a high activity state on 2011 Sept 17. Additionally, we find the X-ray spectra of \pks{} in Paper I to be well accounted for by a single non-thermal power-law, possibly from the jet. However, the relatively sparse sampling makes it difficult to draw strong conclusions regarding a possible correlation.
		
		In previous studies \pks{} has been assigned a flatter spectral index ($\alpha > -0.5$) at low frequencies (<5\,GHz) based on non-simultaneous data (e.g., \oshlack, \citealt{Healey2007}). We calculated the simultaneous spectral index between 1.38\,GHz and 2.4\,GHz from \gallo{} to be $\alpha_{1.38-2.4}=-0.21\pm 0.08$, which is significantly flatter than above 2.4\,GHz ($\alpha=-0.56\pm 0.03$). This suggest that a turnover in the radio spectrum occurs below 2\,GHz as is observed for CSS sources \citep{ODea1998}. 
		
		The weighted average spectral luminosity at $5\,\mathrm{GHz}$ is calculated by extrapolating the flux density of our ATCA data at 5.5\,GHz. We use the fitted spectral index for the $K$-correction following e.g., \cite{Ghisellini2009}. This yields $L_{5,\mathrm{wm}}=(7.4\pm 1.1)\times 10^{25}\mathrm{\,W\,Hz^{-1}}$, which is consistent with the lower end of the 5\,GHz luminosity distribution in the CSS/GPS-sample discussed by \cite{ODea1998}. The combination of archival and our new flux density measurements supports the classification of \pks{} as a CSS source as originally suggested by \oshlack.

	\subsection{Parsec-scale structure and jet orientation}
	\label{subsec-comparison-vlba-image}
		
		The 1.5\,GHz VLBA image shows a bright central component and a jet towards the north-west and substantial diffuse emission towards the east. The central and western structure is consistent with an image based on the same data by \cite{Orienti2015}, in which the diffuse eastern emission, however, is not visible because of the smaller field of view.
		
		We measure high brightness temperatures of $>10^{10}\mathrm{\,K}$ in the VLBI core at both frequencies. The non-simultaneous spectral emission of the core is relatively flat with a spectral index of $\alpha_\mathrm{VLBI,1.5-8.4}\sim -0.04$. A core-jet structure has been observed in some CSS sources but is more common in GPS sources (e.g., \citealt{ODea1998,Fanti2001} and references therein).
		
		The angle of the jet to the line of sight $\theta$ can be estimated using the jet-to-counterjet ratio $R$, the intrinsic jet velocity $\beta$, and the spectral index $\alpha$ via,
		\begin{align}
		R = \frac{S_\mathrm{J}}{S_\mathrm{CJ}} = \left(\frac{1+\beta \cos \theta}{1-\beta\cos\theta}\right)^{2-\alpha}
		\label{eq:angle}
		\end{align}
		where $S_\mathrm{J}$ and $S_\mathrm{CJ}$ are the jet and counter-jet flux densities, respectively (e.g., \citealt{Urry1995}). Assuming that the diffuse emission on the eastern side of the core at 1.5\,GHz can be attributed to the counter-jet, yields $S_\mathrm{CJ}\approx 59\mathrm{\,mJy}$. Excluding the core region results in an integrated flux density in the jet of $S_\mathrm{J}\approx 313\mathrm{\,mJy}$. This leads to an upper limit of $\theta_\mathrm{LOS, 1.5}\lesssim 74\degr$ for an optically thin jet with spectral index of $\sim -1$ \citep{Hovatta2014} and $\beta\rightarrow 1$. As the counter-jet is not detected at 8.4\,GHz, we use the peak flux density per beam in the first feature of the jet next to the core ($S_\mathrm{J} \approx 9.5 \mathrm{\,mJy\,beam}^{-1}$) and five times the noise level on the side of the counter-jet as a detection limit ($S_\mathrm{CJ} \approx 0.1 \mathrm{\,mJy\,beam}^{-1}$). This yields a lower upper limit for the jet angle to the line of sight of $\theta_\mathrm{LOS,8.4}\lesssim 50\degr$. Conversely, the jet-to-counter-jet ratios provide a lower limit on the intrinsic jet velocity, with $\beta_{1.5}>0.27$ and $\beta_{8.4}>0.67$ at 1.5\,GHz and 8.4\,GHz, respectively. Differences in the results between both frequencies can be perhaps attributed at least in part to different scales of the emission region used for the estimate of $S_\mathrm{J}$ and $S_\mathrm{CJ}$. This provides an independent proof of relativistic bulk motion in \pks.
		
		The coherent emission at $\sim 45\mathrm{\,mas}$ may indicate a moderately pronounced `hotspot' with a brightness temperature of $10^7\mathrm{\,K}$ at 1.5\,GHz. Such brightness temperatures have also been measured in `hot spots' of CSS/GPS sources \citep{Dallacasa2002a,Dallacasa2002b}. Diffuse emission north of the `hotspot' can be seen in both images. The offset in position angle may indicate a backflow of jet plasma after an interaction with the surrounding medium. The Eastern diffuse emission is not detected in our TANAMI image at 8.4\,GHz. Given the non-detection on a $5\sigma_\mathrm{RMS}$ level this indicates a spectral index steeper than $-$1.9. The full extent of the radio emission from \pks{} at 1.5\,GHz picked up by the VLBA is $\sim 140\mathrm{\,mas}$, which corresponds to a projected linear size of roughly 530\,pc. 
		
		The linear size of \pks{} can also be assessed indirectly from the radio spectrum. \cite{ODea1997} studied the relationship between rest turnover frequency $\nu_m$ and $l_\mathrm{ps}$ of a sample of CSS and GPS sources, finding an anti-correlation $\nu_m\propto l_{ps}^{-0.65}$. Since the turnover in the radio spectrum is only partially visible below 2\,GHz, we adopt the lowest frequency measurement of the PKS90CAT data as an estimate for the turnover frequency with $\nu_\mathrm{m,\mathrm{obs}}=408\mathrm{\,MHz}$ and $\nu_\mathrm{m}\sim 500\mathrm{\,MHz}$ in the observer and source rest frame, respectively. This yields an estimate of the linear size from the spectrum of $l_\mathrm{ps} \sim 2\mathrm{\,kpc}$ which is likely overestimating the true size, but still consistent with a CSS source.
		
		Another estimate of $l_\mathrm{ps}$ can be obtained from the ATCA Calibrator Database data at 4\,cm. Here, \pks{} shows a flat visibility data distribution (out to $d_\mathrm{max} \approx 150\mathrm{\,k}\lambda$) in the $(u,v)$-plane and no significant defect, which suggests that \pks{} remains unresolved. Using $\sin \phi \approx 1.22 \lambda/d_\mathrm{max}$, this yields an angular size of $\phi\lesssim 1\farcs7$ to a linear projected size of only $l_\mathrm{ps} \lesssim 6.4\mathrm{\,kpc}$, corresponding to $\nu_{m} \gtrsim 240\,\mathrm{MHz}$ and $\nu_\mathrm{m,obs} \gtrsim 200\,\mathrm{MHz}$ for the turnover frequency in the source and observer frame, respectively.
		
		The various estimates of the linear size and the turnover frequencies of \pks{} are all consistent with CSS sources. Nevertheless, we adopt the directly measured limit of $l_\mathrm{ps}\lesssim 11\mathrm{\,kpc}$ from the VLA image as a conservative upper limit on the linear size. Recently, \cite{Richards2015} detected extended emission in three \rlnls{s} using observations with the VLA at a frequency of 9.0\,GHz. The study achieved sub-arcsec resolution and high-sensitivity between $0.021\mathrm{\,mJy/beam}$ and $0.05\mathrm{\,mJy/beam}$. The authors suggested that extended emission will be detected in more sources if higher sensitivity observations would be conducted. The size of the extended emission in these sources is larger than the beam of our VLA observation at 1.5\,GHz, indicating that extended emission may not have been detected due to the sensitivity. In order to further analyse this, we extrapolated the detection limit of our VLA observation at 1.5\,GHz, i.e. three times the noise level ($0.48\mathrm{\,mJy/beam}$) to 9.0\,GHz using a spectral index of $-0.7$ for synchrotron emission. This yields a flux density of $\sim 0.14\mathrm{\,mJy/beam}$ corresponding to 14 times the lowest sensitivity in \cite{Richards2015}. However, our ATCA data at 9.0\,GHz does not indicate extended emission. At this frequency the resolution of ATCA is about $1\farcs7$ and the sensitivity of this observation is approximately $\sim 0.07\mathrm{\,mJy/beam}$ comparable to \cite{Richards2015}. Hence, we cannot completely rule out the existence of very faint extended emission on scales larger than $11\mathrm{\,kpc}$ (VLA limit) or $6.4\mathrm{\,kpc}$ (ATCA limit), but it would most likely require observations at even higher sensitivity than in \cite{Richards2015}. Hence, we consider $l_\mathrm{ps}\lesssim 11\mathrm{\,kpc}$ as a reasonable upper limit. Applying the $\nu_m$-$l_\mathrm{ps}$-relationship yields $\nu_{m,\mathrm{obs}}\gtrsim 140\mathrm{\,MHz}$, which is currently not accessible by our radio data. 
		
		In a recent study \cite{Kunert-Bajraszewska2010} compared the 1.4\,GHz luminosity with the linear size and redshift of a sample of CSS, GPS and low-luminous compact (LLC) objects. Using the flux density measured by \gallo{} at 1.38\,GHz and $\alpha_{1.38-2.4}$ from Sect. \ref{subsec-Discussion-radio-spectrum} yields a luminosity of $L_{1.38} \approx 1.2\times 10^{26}\mathrm{W\,Hz}^{-1}$. This puts \pks{} at the lower end of the luminosity distribution in the combined CSS sample of \cite{Kunert-Bajraszewska2010}. The linear size of \pks{} agrees well with the CSS sample considering even the lowest possible estimate of the linear size from the VLBA image. In addition, \pks{} fits also with the distribution of LLC sources, where it is at the upper end of the luminosity distribution. This is to be expected as one of the selection criteria of the LLC sample is $L_{1.4} \le 10^{26}\mathrm{\,W\,Hz}^{-1}$. The distribution of $L_{1.4}$ with redshift gives a similar result.
		
		Our study confirms the CSS-like nature of \pks{} suggested by \gallo. As such, \pks{} belongs to an even more elusive class of $\gamma$-ray detected AGN than \gnls{} according to the 3LAC \citep{Ackermann2015}: Only one other CSS source has been associated with a $\gamma$-ray counterpart in the 3LAC. 
		
		While CSS/GPS sources have been discussed in the context of young radio sources or frustrated AGNs (e.g., \citealt{ODea1998,Stanghellini2003,Kunert-Bajraszewska2010} and references therein), more recent publications favour the paradigm of young AGNs with respect to the older blazar population (e.g., \citealt{Fanti2011,Randall2011}). Both paradigms are also discussed for \gnls (e.g., \citealt{Abdo2009a,Foschini2015,Berton2015}) and it has been proposed that CSS and \rlnls{} are linked (e.g., \citealt{Komossa2006,Doi2012,Caccianiga2014,Komossa2015,Gu2015}). Recently, \cite{Caccianiga2015} found evidence of significant star formation activity in a sample of \rlnls{s}, favouring the scenario that \rlnls{s} are young AGNs. A similar case is the recently reported $\gamma$-ray loud AGN PMN\,J1603-4904, which also shows a moderately steep radio spectrum \citep{Mueller2014}. It is compact on arcsec-scales and highly symmetric on mas-scales, reminiscent of a compact symmetric object which are also considered to be young radio sources (e.g., \citealt{Wilkinson1994,An2012}).

	\subsection{Comparison with other \gnls}
	\label{sub-sec:comparison}
	
		Among the small sample of \gnls{s} (see \citealt{Foschini2015,Dammando2015,Yao2015}) the sources besides \pks{}, which were studied best so far in the radio regime and which yielded a significant\footnote{$\mathrm{TS}\geq25$, see \cite{Foschini2015,Dammando2015,Yao2015}} $\gamma$-ray detection by \fermi{} are listed in Table \ref{tab-NLSy1-power}. \pks{} is the only \gnls{} detected that is located in the southern hemisphere and it is the radio loudest of them. In this section, we compare our results for \pks{} with these other \gnls{s}. A comparison of the X-ray properties is presented in Paper I.
		
		\subsubsection{Flux density variability and spectrum}
		In Sect. \ref{sec-results} we showed that \pks{} exhibits radio flux density variability up to a factor of 2 on time scales of months to years at centimeter wavelengths, comparable to values reported for other \gnls{s} (e.g., \citealt{Zhou2007, Dammando2013a, Dammando2013b, Angelakis2015}). The recent studies by \cite{Foschini2015} and \cite{Angelakis2015} presented the densest, simultaneous multi-frequency radio monitoring of four \gnls{s}. The strength of the flaring activity differs for each source and is strongly frequency dependent with a frequency delay of the flares as previously known for blazars. 
		To compare the variability index $V_\nu$ calculated for \pks{} at 5.5\,GHz and 9.0\,GHz in Sect. \ref{subsec-atca}, we calculate $V_\nu$ for the \gnls{s} discussed by \cite{Angelakis2015} and list the results for $\sim 5\mathrm{\,GHz}$ and $\sim 8.4\mathrm{\,GHz}$ in Table \ref{tab-NLSy1-power}. \pmn{} has the highest variability index of the five sources, which is also the case at other frequencies except for 23\,GHz and 32\,GHz which is dominated by \h{} despite the lower sampling of \h. \sbs{} has the lowest cadence of the four sources in \cite{Angelakis2015}, though similar to \pks. A comparison with the distribution of $V_\nu$ at the two frequencies of the sample in \cite{Hovatta2008}, which consisted mostly of blazars, showed that the \gnls{s} are located at the lowest end of the distribution except for \pmn, which is consistent with the median values for quasars and BL\,Lacs.
		
		The study by \cite{Angelakis2015} revealed strong differences in the simultaneous radio spectra which resemble types of radio spectra found in blazars by the F-GAMMA program (see \citealt{Angelakis2012}), supporting \cite{Yuan2008} regarding the link between \rlnls{} and blazars using simultaneous data. However, a CSS-like spectrum of \pks{}, with a possible turnover in the MHz-range, does not fit into the classification by \cite{Angelakis2012} and is unique among all \gnls{s}. \cite{Berton2015} compared the population of flat-spectrum \nls{s} from \cite{Foschini2015}, which includes all \gnls{s} with a sample of steep-spectrum \nls{s}. Based on the Kolmogorov-Smirnov test, their findings suggest that both samples may stem from the same population. \cite{Foschini2015} interpret this in such a way that steep-spectrum \nls{s} have a larger angle to the line of sight than flat-spectrum \nls{s}. These characteristics fit to \pks{} with the difference that no steep-spectum \nls{} has been detected in the $\gamma$-ray regime so far. The single exception may be RX\,J2314.9+2243, for which the $\gamma$-ray detection is only tentative (\citealt{Komossa2015} and references therein).
		
		The variety in the radio spectra of \gnls{s} is not matched in the X-rays (see Paper I and \citealt{Foschini2015}), but differences seem to exist in the flaring activity of these sources. For this purpose, we calculated the variability index based on X-ray data $V_X$ from \cite{Dammando2013a,Dammando2013b} and \cite{Foschini2015} for the northern sources and Paper I for \pks. The amount of available data strongly varies between the four sources. \pkss{} has the lowest amount of data, which covers the shortest time range. All of its flux measurements were consistent with each other within the uncertainties. For \h, \pmn{}, and \pks{} the sampling is sufficient for comparison with $V_\nu$. Again, \pmn{} yields the highest value, but \sbs{} and \pks{} show $V_X > V_\nu$ suggesting stronger variability at X-ray energies. 
		The photon index of the X-ray spectra of the \gnls{s} lies between 1.5 and 2.0 (Paper I), which fits into the range covered by BL\,Lac objects \citep{Foschini2015}. However, the comparison of the $\gamma$-ray and X-ray luminosity by \cite{Foschini2015} indicated a distribution similar to flat spectrum radio quasars.
		
		\subsubsection{Radio-morphology and brightness temperature}
		Our deep VLBI observation of \pks{} confirms that on parsec-scales, all considered \gnls{s} possess a one-sided jet extending from a dominant central core (e.g., \citealt{Doi2006,Dammando2013a,Dammando2013b,Wajima2014}). This morphology is reminiscent of blazars (e.g., \citealt{Lister2013} and references therein). In contrast to this, the kilo-parsec scale projected linear size is usually smaller compared to radio-loud AGNs (e.g., \citealt{Yuan2008,Doi2012} and references therein) and the extent of the large scale structure has been measured directly for only four sources (see Table \ref{tab-NLSy1-power}). 
		
		The one-sidedness of the radio jet is a strong indication of relativistic beaming, which often gives rise to apparent superluminal motion of distinct jet features (see e.g., \citealt{Lister2013}). \sbs{} is the only source in this group, for which apparent superluminal motion of $\beta_\mathrm{app}=(9.3\pm 0.6)c$ \citep{Dammando2013b} has been reported so far. However, \cite{Lister2013} reported a maximum apparent speed of 0.82\,c for \hb.
		
		Another indicator of relativistic beaming is the brightness temperature determined from the VLBI core $T_\mathrm{B}$. We computed the mean $T_\mathrm{B}$-value at 15\,GHz and $\sim$8\,GHz for the northern \gnls{s} based on available data from \cite{Doi2012}, \cite{Lister2013}, and \cite{Foschini2015}. In case of \hb, the $T_\mathrm{B}$-value is directly taken from \cite{Pushkarev2012}. We find $T_\mathrm{B}$ to be well in excess of the equipartition limit and in three cases above the inverse-Compton limit. Such values are consistent with blazars (e.g., \citealt{Kovalev2005}). \pks{} shows the lowest $T_B$-value in the sample, about two orders of magnitude below the highest observed value by \hb. This suggests intrinsic differences in Doppler boosting of the jet emission, which is consistent with studies of the variability brightness temperature $T_\mathrm{B,var}$ from radio light curves (e.g., \citealt{Angelakis2015}). 
		
		\begin{table*}
			\caption{Radio properties of \gnls}
			\label{tab-NLSy1-power}
			\centering
			\begin{tabular}{@{\;}c@{\;\;\;}c@{\;\;\;}c@{\;\;\;}c@{\;\;\;}c@{\;\;\;}c@{\;\;\;}c@{\;\;\;}c@{\;\;\;}c@{\;\;\;}c@{\;\;\;}c@{\;}}
				\hline \hline	
				Source & z\tablefootmark{a} & $T_{b,\mathrm{15}}$\tablefootmark{b} & $T_{b,\mathrm{8}}$\tablefootmark{c} & $l_\mathrm{ps}$\tablefootmark{d} & radio spectrum\tablefootmark{e} & variability\tablefootmark{f} & $V_{4.85/5.5}$\tablefootmark{g} &  $V_{8.35/9.0}$\tablefootmark{g} & $V_{X}$\tablefootmark{h} & RL$_{1.4}^{i}$\\
				& 	& [$10^{11}$\,K] & [$10^{11}$\,K] & [kpc] & & &  &  & &\\
				\hline
				\h & 0.061 &  $7.3\pm 2.7$  & - & 24 & steep, flat & F/S & 0.30 & 0.32 & 0.37 & 318\\ 
				&	& $(6.2)$  &  &	&	&  & & & & \\ 
				\sbs &  0.584 & $36\pm 27$ & - & <2.2 & flat, inverted & F/S &  0.13 & 0.14 & 0.42 & 4496\\
				&	&	$(7.5)$	&	&	&	& & & & & \\
				\pmn & 0.585 & $64\pm15$  & $4.0\pm 1.3$ & 104 & flat, inverted & F/S & 0.55 & 0.65 & 0.50 & 846\\
				&	& $(44)$	&	$(1.7)$	&	&	& & & & & \\
				\hb & 0.966 & $119\pm 95$ & 43 & $\sim 80$ & flat & F/- & - & - & - & 1700\\
				&  & (96) & & & & & & & & \\
				\pkss & 0.409 & $3.9\pm 1.1$ & - & <25 & GHz-peaked, flat & F/S & 0.27 & 0.30 & -0.4 & 3364\\
				&	& $(3.0)$ & &  &	& & & & & \\
				\fbqs & 0.145 & - & 6.3 & 60 & - & - & - & - & - & 396 \\
				\pks & 0.24 & -  & $0.4\pm 0.1$  & <11 & CSS-like, steep & F/S & 0.23 & 0.32 & 0.45 & 6358 \\
				\hline
			\end{tabular}
			\tablefoot{Hyphens indicate that no data were available.\\
				\tablefoottext{a} {Redshift from NED;}\\
				\tablefoottext{b} {Mean brightness temperature $T_b$ at 15\,GHz based on (1) except for \hb (2), brackets refer to the median;}\\
				\tablefoottext{c} {$T_b$ at $\sim$\,8\,GHz, mean $T_b$ based on (1) and the median in brackets for \pmn; \h{} from (3), \fbqs{} from (4), \hb{} from (5);}\\
				\tablefoottext{d} {Projected linear size from VLA observations; for \sbs{} and \pkss{} the major beam axis of the highest resolution VLA observation from (6) and (7), respectively, was used as a conservative upper limit; Reference (4) for \h, \pmn, \fbqs; Reference (8) for \hb}\\ 
				\tablefoottext{e} {References for \h{} (3, 9), \sbs{} (6, 9, 10), \pmn{} (11, 9), \pkss{} (7, 9), \fbqs{} (4);}\\ 
				\tablefoottext{f} {F: flux variability; S: spectral variability from simultaneous data;}\\
				\tablefoottext{g} {Variability index for \h, \sbs, \pmn{} and \pkss{} at 4.85 and 8.35\,GHz, respectively, from (9); \pks{} at 5.5 and 9.0\,GHz;}\\
				\tablefoottext{h} {Variability index at X-ray energies from 0.3--10\,keV for \h{} (1), \sbs{} (10), \pmn{} (1), and \pkss{} (7); 0.5--10\,keV for \pks{} (12); For \pkss{} all flux measurements were consistent within the given uncertainties, yielding a negative value;}\\
				\tablefoottext{i} {Reference (13) for all except \hb{} (14);}\\
				References: (1) \cite{Foschini2015}, (2) \cite{Lister2013}, (3) \cite{Wajima2014}, (4) \cite{Doi2012}, (5) \cite{Pushkarev2012}, (6) \cite{Dammando2012}, (7) \cite{Dammando2013a}, (8) \cite{Kharb2010}, (9) \cite{Angelakis2015}, (10) \cite{Dammando2013b}, (11) \cite{Dammando2014}, (12) Paper I, (13) \cite{Foschini2011}, (14) \cite{Yao2015}
			}
		\end{table*}

\section{Summary and Conclusion}
\label{sec-sum}

	We have presented the highest resolution VLBI image so far obtained for the \gnls{} galaxy \pks. \pks{} is the only southern source in an elusive sample and the radio-loudest one. The image reveals a single-sided jet extending to the north-west from a dominant core with a brightness temperature of $(5.7\pm 1.4)\times 10^{10}\mathrm{\,K}$. The first long-term multi-frequency flux density measurements with ATCA between 5.5\,GHz and 40\,GHz demonstrates a persistent steep radio spectrum with a spectral index betweeen $-0.5$ and $-0.9$ and moderate flux density variability. 
	
	We compared our results within the small sample of \gnls{s}. We find that the brightness temperature of the VLBI cores are either close to or in excess of the equipartition limit ranging over two orders of magnitude. In combination with the single-sided jet morphology on parsec-scales this demonstrates the importance of relativistic beaming of the radio emission in all of these sources. In addition, the variability index at $\sim 5\mathrm{\,GHz}$ and $\sim 8\mathrm{\,GHz}$ shows significant differences in variability. The shape of the radio spectrum differs strongly among \gnls{s} with the persistent CSS-like spectrum of \pks{} being a unique exemplar. It is also extremely rare among $\gamma$-ray loud AGN listed in the 3LAC.
	
	In paper I, we showed that the X-ray spectra of \gnls{s} do not show as much diversity as the radio spectra. In this paper, we calculated the variability index at X-ray energies finding that \pmn{} has the highest variability among sources with sufficient data.
	
	The TANAMI program continues the VLBI observations and flux density measurements of \pks{} to study temporal changes in the jet and radio spectrum to constrain or measure the jet speed. An accurate measurement of the turnover frequency would come from observations below $400\mathrm{\,MHz}$, making \pks{} an intriguing candidate for studies with high sensitivity, low-frequency precursors of the Square Kilometre Array (e.g., \citealt{Wayth2015}).

\begin{acknowledgements}
We thank the MPIfR internal referee S. Komossa and the anonymous journal referee for their insightful comments, which improved the manuscript in its final form.
This work was funded by Deutsche Forschungsgemeinschaft grant WI 1860/10-1. 
We acknowledge support by the Deutsches Zentrum f\"{u}r Luft- und Raumfahrt under contract number 50 OR 1303, the Spanish MINECO projects AYA2009-13036-C02-02 and AYA2012-38491-C02-01 and by the Generalitat Valenciana projects PROMETEO/2009/104 and PROMETEO/II/2014/057, as well as by the MP0905 action 'Black Holes in a Violent Universe'. 
This research was funded in part by NASA through Fermi Guest Investigator grants NNH09ZDA001N, NNH10ZDA001N, NNH12ZDA001N, and NNH13ZDA001N-FERMI (proposal numbers 31263, 41213, 61089, and 71326 respectively). 
This research was supported by an appointment to the NASA Postdoctoral Program at the Goddard Space Flight Center, administered by Oak Ridge Associated Universities through a contract with NASA.
The Australian Long Baseline Array and the Australia Telescope Compact Array are part of the Australia
Telescope National Facility which is funded by the Commonwealth of Australia for operation as a National Facility managed by CSIRO.
This research has made use of the Interactive Spectral Interpretation System (ISIS) \citep{Houck2000}. This research has made use of a collection of ISIS scripts provided by the Dr. Karl Remeis observatory, Bamberg, Germany at http://www.sternwarte.uni-erlangen.de/isis/. This research has made use of the NASA/IPAC Extragalactic Database (NED) which is operated by the Jet Propulsion Laboratory, California Institute of Technology, under contract with the National Aeronautics and Space Administration. This research has made use of the VizieR catalogue access tool, CDS, Strasbourg, France.
This research has made use of data from the MOJAVE database that is maintained by the MOJAVE team (Lister et al., 2009, AJ, 137, 3718).
\end{acknowledgements}

\bibliographystyle{aa} 
\bibliography{References} 


\begin{appendix}
	
	\section{Data from ATCA Observations}
	\label{app:Data}
	
	\begin{table}
		\caption{Details from ATCA Observations}
		\label{tab-AtcaObsInfo}
		\centering
		\begin{tabular}{rcc}
			\hline\hline
			$\nu$ & Date & $S_\nu$ \\
			$[$GHz$]$ & [YYYY-MM-DD] & [Jy] \\
			\hline
			5.5 & 2010-02-13 & $0.52\pm0.02$ \\
			9.0 & 2010-02-13 & $0.38\pm0.02$ \\
			43.0 & 2010-05-09 & $0.16\pm0.02$ \\
			45.0 & 2010-05-09 & $0.15\pm0.02$ \\
			1.7 & 2010-07-03 & $0.77\pm0.03$ \\
			35.0 & 2010-07-26 & $0.18\pm0.02$ \\
			33.0 & 2010-07-26 & $0.19\pm0.02$ \\
			43.0 & 2010-09-14 & $0.10\pm0.01$ \\
			45.0 & 2010-09-14 & $0.10\pm0.01$ \\
			2.1 & 2010-12-17 & $0.67\pm0.03$ \\
			33.0 & 2011-04-12 & $0.16\pm0.02$ \\
			35.0 & 2011-04-12 & $0.16\pm0.02$ \\
			5.5 & 2011-05-17 & $0.41\pm0.02$ \\
			9.0 & 2011-05-17 & $0.29\pm0.01$ \\
			17.0 & 2011-05-17 & $0.18\pm0.01$ \\
			19.0 & 2011-05-17 & $0.17\pm0.01$ \\
			5.5 & 2011-10-14 & $0.41\pm0.02$ \\
			9.0 & 2011-10-14 & $0.29\pm0.01$ \\
			17.0 & 2011-10-14 & $0.18\pm0.01$ \\
			19.0 & 2011-10-14 & $0.16\pm0.01$ \\
			5.5 & 2011-11-08 & $0.40\pm0.02$ \\
			9.0 & 2011-11-08 & $0.26\pm0.01$ \\
			17.0 & 2011-11-08 & $0.18\pm0.01$ \\
			19.0 & 2011-11-08 & $0.16\pm0.01$ \\
			38.0 & 2011-11-08 & $0.13\pm0.01$ \\
			40.0 & 2011-11-08 & $0.13\pm0.01$ \\
			5.5 & 2011-11-27 & $0.37\pm0.01$ \\
			9.0 & 2011-11-27 & $0.28\pm0.01$ \\
			17.0 & 2011-11-27 & $0.20\pm0.01$ \\
			19.0 & 2011-11-27 & $0.18\pm0.01$ \\
			5.5 & 2012-01-15 & $0.38\pm0.02$ \\
			9.0 & 2012-01-15 & $0.25\pm0.01$ \\
			5.5 & 2012-03-17 & $0.41\pm0.02$ \\
			9.0 & 2012-03-17 & $0.27\pm0.01$ \\
			5.5 & 2012-04-22 & $0.33\pm0.01$ \\
			9.0 & 2012-04-22 & $0.19\pm0.01$ \\
			17.0 & 2012-05-28 & $0.15\pm0.01$ \\
			19.0 & 2012-05-28 & $0.14\pm0.01$ \\
			38.0 & 2012-05-28 & $0.09\pm0.01$ \\
			40.0 & 2012-05-28 & $0.09\pm0.01$ \\
			5.5 & 2012-06-13 & $0.41\pm0.02$ \\
			9.0 & 2012-06-13 & $0.29\pm0.01$ \\
			17.0 & 2012-06-13 & $0.16\pm0.01$ \\
			19.0 & 2012-06-13 & $0.14\pm0.01$ \\
			38.0 & 2012-06-13 & $0.08\pm0.01$ \\
			40.0 & 2012-06-13 & $0.08\pm0.01$ \\
			17.0 & 2012-09-07 & $0.19\pm0.01$ \\
			19.0 & 2012-09-07 & $0.17\pm0.01$ \\
			5.5 & 2012-11-02 & $0.45\pm0.02$ \\
			9.0 & 2012-11-02 & $0.33\pm0.01$ \\
			5.5 & 2013-11-10 & $0.44\pm0.02$ \\
			9.0 & 2013-11-10 & $0.31\pm0.01$ \\
			5.5 & 2014-03-25 & $0.49\pm0.02$ \\
			9.0 & 2014-03-25 & $0.30\pm0.01$ \\
			5.5 & 2014-03-26 & $0.56\pm0.02$ \\
			9.0 & 2014-03-26 & $0.41\pm0.02$ \\
			17.0 & 2014-03-26 & $0.28\pm0.01$ \\
			19.0 & 2014-03-26 & $0.26\pm0.01$ \\
			\hline
		\end{tabular}
	\end{table}
	
\end{appendix}
\label{sec-appendix}
\clearpage
\appendix

\end{document}